\UseRawInputEncoding
\documentclass[aps,prd,twocolumn,groupedaddress]{revtex4-2}

\usepackage{latexsym}
\usepackage{graphics}
\usepackage{graphicx}
\usepackage{epstopdf}
\usepackage{epsfig}
\usepackage{amssymb}
\usepackage{makeidx}
\usepackage{nicefrac}
\usepackage{amsmath}

\begin{document}


\title{Properties of Liquid Argon Scintillation Light Emission}


\author{Ettore Segreto}
\email[]{segreto@ifi.unicamp.br}
\affiliation{Instituto de F\'isica ``Gleb Wataghin'' Universidade Estadual de Campinas - UNICAMP\\ Rua S\'{e}rgio Buarque de Holanda, No 777, CEP 13083-859 Campinas, S\~ao Paulo, Brazil}


\date{\today}

\begin{abstract}
Liquid argon is used as active medium in a variety of neutrino and Dark Matter experiments thanks to its excellent properties of charge yield and transport and as a scintillator. Liquid argon scintillation photons are emitted in a narrow band of 10~nm centered around 127~nm and with a characteristic time profile made by two components originated by the decay of the lowest lying singlet, $^1\Sigma_u^+$, and triplet states, $^3\Sigma_u^+$,  of the excimer Ar$_2^*$ to the dissociative ground state. A model is proposed which takes into account the quenching of the long lived triplet states through the self-interaction with other triplet states or through the interaction with molecular Ar$_2^+$ ions. The model predicts the time profile of the scintillation signals and its dependence on the intensity of an external electric field and on the density of deposited energy, if the relative abundance of the unquenched fast and slow components is know. The model successfully explains the experimentally observed dependence of the characteristic time of the slow component on the intensity of the applied electric field and the increase of photon yield of liquid argon when doped with small quantities of xenon (at the ppm level). The model also predicts the dependence of the pulse shape parameter, F$_{prompt}$, for electron and nuclear recoils on the recoil energy and the behavior of the relative light yield of nuclear recoils in liquid argon, $\mathcal{L}_{eff}$.  
\end{abstract}


\maketitle

\section{Introduction\label{intro}}
Liquid Argon (LAr) is a powerful medium to detect ionizing particles and is widely used in neutrino and Dark Matter experiments since several years 
\cite{Abi_2020}, \cite{Rubbia-2011}, \cite{WArP}, \cite{AGNES2015456-darkside50}, \cite{Chen:2007ae-microboone}, \cite{SBND-Ornella}.
LAr scintillation photons are emitted in the Vacuum Ultra Violet in a 10~nm band centered around 127~nm  with a time profile made by two components with very different characteristic decay times, a fast one in the nanosecond range and a slower one in the microsecond range \cite{Doke:1988rq}. The relative abundance of the two components depends strongly on the ionizing particle type and allows for a powerful particle discrimination \cite{ACCIARRI20121113} \cite{Wahl_2014}.
LAr scintillation has been deeply studied by several authors \cite{Doke:1988rq}, \cite{PhysRevB.46.11463}, \cite{Hitachi2019PropertiesFL} and a solid understanding of the main mechanisms regulating the production, emission and propagation of scintillation photons has been achieved. 
However, there are experimental results which can not be easily explained and described by the currently accepted models,
as the dependence of the characteristic time of the slow component on the intensity of the electric field applied to LAr and the increase of the LAr photon yield when doped with small quantities of xenon, at the level of few ppm (part per million). These phenomena point to quenching mechanisms involving the excited species which are the precursors of the scintillation photons and which have not been investigated before.\\
The model proposed in this work  takes into account these quenching processes and
predicts the shape of the scintillation pulse as a function of the applied electric field and of the density of the energy transferred to the electrons of LAr.  
In addition to the experimental observations mentioned above, it also allows to explain the dependence of the pulse shape parameter, F$_{prompt}$, on the energy of the incoming particle. The cases of electron and nuclear recoils are explicitly treated and compared to data.\\
The integral of the scintillation pulse allows to estimate the amount of quenching due to these processes and to predict the behavior of the relative light yield of nuclear recoils in LAr,  $\mathcal{L}_{eff}$, as a function of the recoil energy, which is a quantity of central importance for Dark Matter experiments.


\section{LAr scintillation Model\label{model}}
The passage of ionizing particles in LAr produces free excitons and electron-hole pairs. The proportion between these two species is assumed to be independent of the ionizing particle type and energy: N$_{ex}$/N{$_i$}~=~0,21, where N$_{ex}$ is the abundance of excitons and N$_i$ the abundance of electron-hole pairs. Free excitons and holes are self-trapped within about 1~ps from their production and result into excited, Ar$_2^*$, or ionized, Ar$_2^+$, argon dimers. Ar$_2^+$ recombines with a thermalized electron to form Ar$_2^*$ \cite{PhysRevB.46.11463} which in turn decays non-radiatively to the first singlet and triplet excited states $^1\Sigma_u^+$ and $^3\Sigma_u^+$. These two states, whose dis-excitation leads to the emission of the scintillation photons, have approximately the same energy with respect to the dissociative ground state, while the lifetimes are very different: in the nanosecond range for $^1\Sigma_u^+$ and in the microsecond range for $^3\Sigma_u^+$ \cite{Doke:1988rq}.\\
The scintillation photon yield of LAr
depends on the ionizing particle type and on the Linear Energy Transfer (LET) \cite{DOKE1985136}. The highest photon yield is reached by relativistic heavy nuclei, from Ne to La. Low LET light particles (e$^-$, p) have a slightly reduced photon yield due to the fact that a fraction of the ionization electrons escapes from recombination. Nuclear recoils and $\alpha$ particles also have a reduced photon yield, but the quenching mechanism is different and not fully clarified yet. A bi-excitonic interaction in the core of the track has been proposed  as a possible explanation \cite{PhysRevB.46.11463} and a reasonably good agreement is found for 5,3 MeV $\alpha$ particles, while the model is not accurate in explaining the dependence of the photon yield for nuclear recoils \cite{Hitachi2019PropertiesFL} at very low energies ($\sim$ tens of keV). In this case, a better agreement with the available data is obtained when using the Birks law to account for the possible quenching processes involving excitons and excited dimers Ar$_2^*$ \cite{Mei2008AMO} \cite{Birks_1951}.\\
    
Two recent experimental observations can not be easily explained with the current understanding of the LAr scintillation process.
The DUNE (Deep Underground Neutrino Experiment) Collaboration has reported a clear dependence of the lifetime of the triplet state $^3\Sigma_u^+$ of the $Ar_2^*$ dimer on the electric field in which the LAr is immersed \cite{Lastoria:2019qiw}.
The scintillation light was produced by a sample of  cosmic muons crossing one of the prototypes of the dual phase DUNE detector, the 4-ton demonstrator \cite{Aimard_2018}. The light was detected with an array of five 8" photomultipliers (Hamamatsu R5912-02Mod) coated with a wavelength shifter, TetraPhenyl-Butadiene (TPB) \cite{PhysRevC.91.035503}, to convert the 127 nm photons to 430 nm and the electric field was varied between 0 and 600 V/cm.\\
The second experimental evidence is related to the doping of LAr with small concentrations of xenon.  It has been reported that adding a few tens of ppm of xenon to LAr has the effect of shifting the wavelength of the triplet component from 127~nm to 174~nm, shortening the signal from few $\mu$sec to hundreds of nsec  and enhancing the Light Yield (LY) \cite{Wahl_2014}. The enhancement of LY can not be explained by an higher quantum efficiency of the wavelength shifter (TPB) for 174~nm than for 127~nm, since it has been measured to be almost the same \cite{TPB_efficiency} and should be attributed to an increase of the LAr photon yield.\\
These two effects point to quenching processes of the triplet states and to an hidden amount of light which has not been described before.
The proposed quenching mechanisms for the triplet states are two: one relies on the interaction of two excited dimers Ar$_2^*$ and the other on the interaction of one ionized dimer Ar$_2^+$ with one excited dimer Ar$_2^*$: 
\begin{eqnarray}
\label{eq:quenching_self}
Ar_2^*~+~Ar_2^*\rightarrow Ar_2^*~+~2Ar\\
\label{eq:quenching_ions}
Ar_2^*~+~Ar_2^+\rightarrow Ar_2^+~+~2Ar
\end{eqnarray}
In the absence of an external electric field, reaction \ref{eq:quenching_ions} is possible when escaping electrons prevent the complete recombination of the ionization charge and thus only for low LET particles (electrons, muons, protons, $\dots$).\\ 
It is assumed that only the excited dimers in the triplet state participate to this quenching processes, since the lifetime of the singlet state is too short.
The instantaneous variation of the number density of triplet states, N$_3$, and of the number density of ionized dimers (N$^+$) can be written as:
\begin{eqnarray}
\label{eq:N_3_t_X}
\frac{dN_3}{dt} = D~\nabla^2N_3 -\lambda_3 N_3-\sigma^+ v^+ N^+ N_3 - \sigma_3 v_3N_3^2\\
\label{eq:N+_t_x}
\frac{dN^+}{dt} = D^+~\nabla^2N^+
\end{eqnarray} 
where N$_3$ and N$^+$ depend on time and position in space, D and D$^+$ are the diffusion constants of $Ar_2^*$ and $Ar_2^+$ respectively,   $\lambda_3$ is the radiative dis-excitation rate of the $^3\Sigma_u^+$ state, $\sigma^+$ is the cross section for the process \ref{eq:quenching_ions}, $v^+$ is the relative velocity between a triplet excimer and a ion, 
$\sigma_3$ is the cross section for process \ref{eq:quenching_self} and $v_3$ is the relative velocity of two $Ar_2^*$.\\

In the hypotheses that the diffusion terms can be neglected and that
$Ar_2^*$ and $Ar_2^+$
are uniformly distributed inside a cylinder of radius r$_3$ along the track, equations \ref{eq:N_3_t_X} and \ref{eq:N+_t_x} reduce to one, which depends only on time:
\begin{equation}
\label{eq:tau_triplet}
\frac{dN_3}{dt} = -\lambda_3 N_3-\sigma^+ v^+ N^+_0 N_3 - \sigma_3 v_3N_3^2
\end{equation}
where $N_0^+$ is the density of $Ar_2^+$, which is a constant. The hypotheses of the model will be discussed in section \ref{sec:discussion}.
Equation \ref{eq:tau_triplet} can be solved analytically and gives:
\begin{equation}
N_3(t) = N_0\frac{e^{-\lambda_q t}}{1+q/\lambda_q (1-e^{-\lambda_q t})}
\end{equation}
where $N_0$ is the initial density of triplet states, $\lambda_q = \lambda_3 + k^+$, $k^+ = \sigma^+ v^+ N_0^+$, $q=N_0 \sigma_3 v_3$.
The probability density of the LAr scintillation light can be written as:
\begin{equation}
\label{eq:l_t}
l(t) = \frac{\alpha_s}{\tau_s}e^{-\frac{t}{\tau_s}}+\frac{\alpha_3}{\tau_3}\frac{e^{-\frac{t}{\tau_q}}}{1+q \tau_q (1-e^{-\frac{t}{\tau_q}})}
\end{equation}
where $\alpha_s$ is the initial abundance of the singlet states, $\tau_s$ is the decay time of the singlet states, $\alpha_3$ is the initial abundance of triplet states, $\tau_3$ is the unquenched decay time of the triplet states, and $\tau_q = 1/(\lambda_3 + k^+)$=1/$\lambda_q$. The probability density of triplet states depends on the electric field and on the LET thorough $\tau_q$ and q. The integral, L, of the probability density $l(t)$, which is proportional to the total number of scintillation photons emitted, is given by:
\begin{equation}
\label{eq:light_partition}
L = L_s + L_3 = \alpha_s + \alpha_3~\frac{\ln(1+q~\tau_q)}{q~\tau_3}
\end{equation} 
L is equal to one ($\alpha_s$+$\alpha_3$) only when q is zero and $\tau_q = \tau_3$.
\section{Extraction of the parameters of the model for electron and nuclear recoils}
\label{sec:electron_neutrons}
\begin{figure}
	\includegraphics[width=10cm]{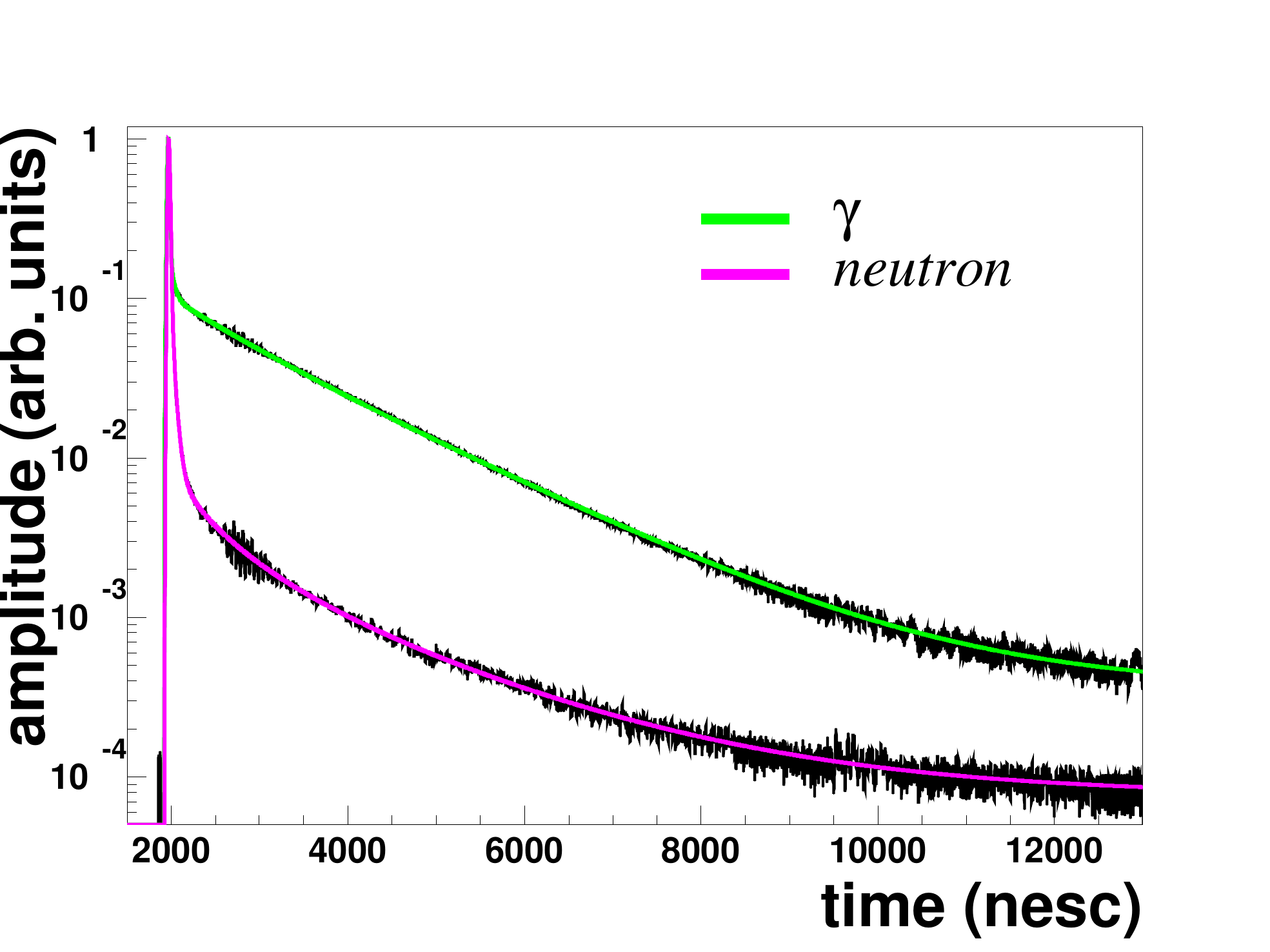}
	\caption{\label{fig:gamma-neutrons} Average waveforms for gammas and neutrons. Green and magenta lines represent the result of the fitting procedure for gammas and neutrons respectively.}
\end{figure}

Some of the parameters of equation \ref{eq:l_t} for electron and nuclear recoils have been extracted through a fit procedure of experimental waveforms. The data which have been analyzed were collected during the test described in \cite{ACCIARRI20121113}, \cite{tesi_rob}. Within the R\&D program of the WArP experiment \cite{WArP} a 4 liters single phase LAr chamber, observed by seven 2" photomultipliers (ETL D749U), was exposed to neutron (AmBe) and $\gamma$ sources. The internal surfaces of the LAr chamber were coated with TetraPhenyl-Butadiene (TPB) \cite{Francini_2013} to down-convert the 127~nm LAr scintillation photons to 430~nm making them detectable by the photomultipliers.    
After a selection of the events based on the shape of the signals, electron and nuclear recoil average waveforms were calculated for different intervals of deposited energy. The LY of the detector was measured to be 1,52~phel/keV (photo-electrons/keV) for the considered run.
An average electron recoil waveform, calculated with signals containing between 130 and 150 phel, and an average nuclear recoil waveform, calculated with signals containing between 150 and 180 phel, have been simultaneously fitted. Considering the LY of the detector these correspond to an energy interval of 85 to 100~keV and of 220 to 290~keV respectively. The fit function contains the amplitudes of the singlet and triplet components ($\alpha_s$ and $\alpha_3$), the decay time of the singlet component ($\tau_s$), the unquenched decay time of the triplet component ($\tau_3$) and the rate constants $k^+$ and $q$. An additional time component, with a decay time around 50~nsec, is included to take into account the late light re-emission of TPB \cite{PhysRevC.91.035503} with a constant abundance with respect to the singlet component. The light signal is convoluted with a Gaussian function to accomodate the statistical fluctuations and the response of the read-out electronics.\\
The parameters which are assumed to be common to the electron and nuclear recoil waveforms are the decay time of the singlet component ($\tau_s$), the unquenched decay time of the triplet component ($\tau_3$) and the fraction of TPB late light, while the other parameters are assumed to be particle dependent. The result of the fit is shown in figure \ref{fig:gamma-neutrons}.        
The main parameters of the fit are reported in table \ref{tab:fit-gamma-neutron}. 

\begin{table}
	\caption{\label{tab:fit-gamma-neutron} Parameters of the model (equation \ref{eq:l_t}) extracted from the fit procedure. The electron recoil average waveform is constructed with events with a deposited energy between 85 and 100~keV, while the nuclear recoil average waveform with events with a deposited energy between 220 and 290~keV.}
		\begin{ruledtabular}
		\begin{tabular}{l|cc}
			  &	gamma & neutron\\
		\hline
		$\tau_3$ (nsec) & 2100 $\pm$ 20 & 2100 $\pm$ 20\\
		$\tau_s$ (nsec) & 5 $\pm$ 5 & 5 $\pm$ 5\\   		
		$\alpha_s$ & 0.14 $\pm$ 0.02 & 0.64 $\pm$ 0.02 \\
		$\alpha_3$ & 0.86 $\pm$ 0.02 & 0.36 $\pm$  0.02\\
		k$^+$ (nsec$^{-1}$)& (1.3$\pm$0.1)$\times$10$^{-4}$ & 0.\\
		q (nsec$^{-1}$)& (2.3$\pm$0.2) $\times$10$^{-4}$ & (2.3$\pm$0.2)$\times$10$^{-3}$ \\  
		\end{tabular}
		\end{ruledtabular}
\end{table}

\section{Dependence of the slow decay time from the electric field at low LET}
The shape of the LAr scintillation waveform depends on the module of the applied electric field, $\mathcal{E}$, through the parameters k$⁺$ and q (see equation \ref{eq:l_t}). The charge recombination factor, R($\mathcal{E}$), is assumed to have the form:
\begin{equation}
\label{eq:recomb_factor}
R(E) = B + \frac{A}{1+ \nicefrac{K_{\mathcal{E}}}{\mathcal{E}}}
\end{equation}
where B takes into account the fraction of the charge which does not recombine even at null electric field, due to escaping electrons, k$_{\mathcal{E}}$ is the Birks recombination constant and A is a normalization constant \cite{AMORUSO2004275}.

Equation \ref{eq:recomb_factor} can be used to make explicit the dependence of the density of Ar$_2^+$ ions, N$_0^+$, and of the initial density of triplet states, N$_0$ on the electric field: 
\begin{eqnarray}
\label{eq:N0+}
N_0^+ = N_i~R(\mathcal{E})\\
\label{eq:q}
N_0 = N_i~\alpha_3~[1-R(\mathcal{E})+\nicefrac{N_{ex}}{N_{i}}]
\end{eqnarray} 

\begin{figure}
	\includegraphics[width=10cm]{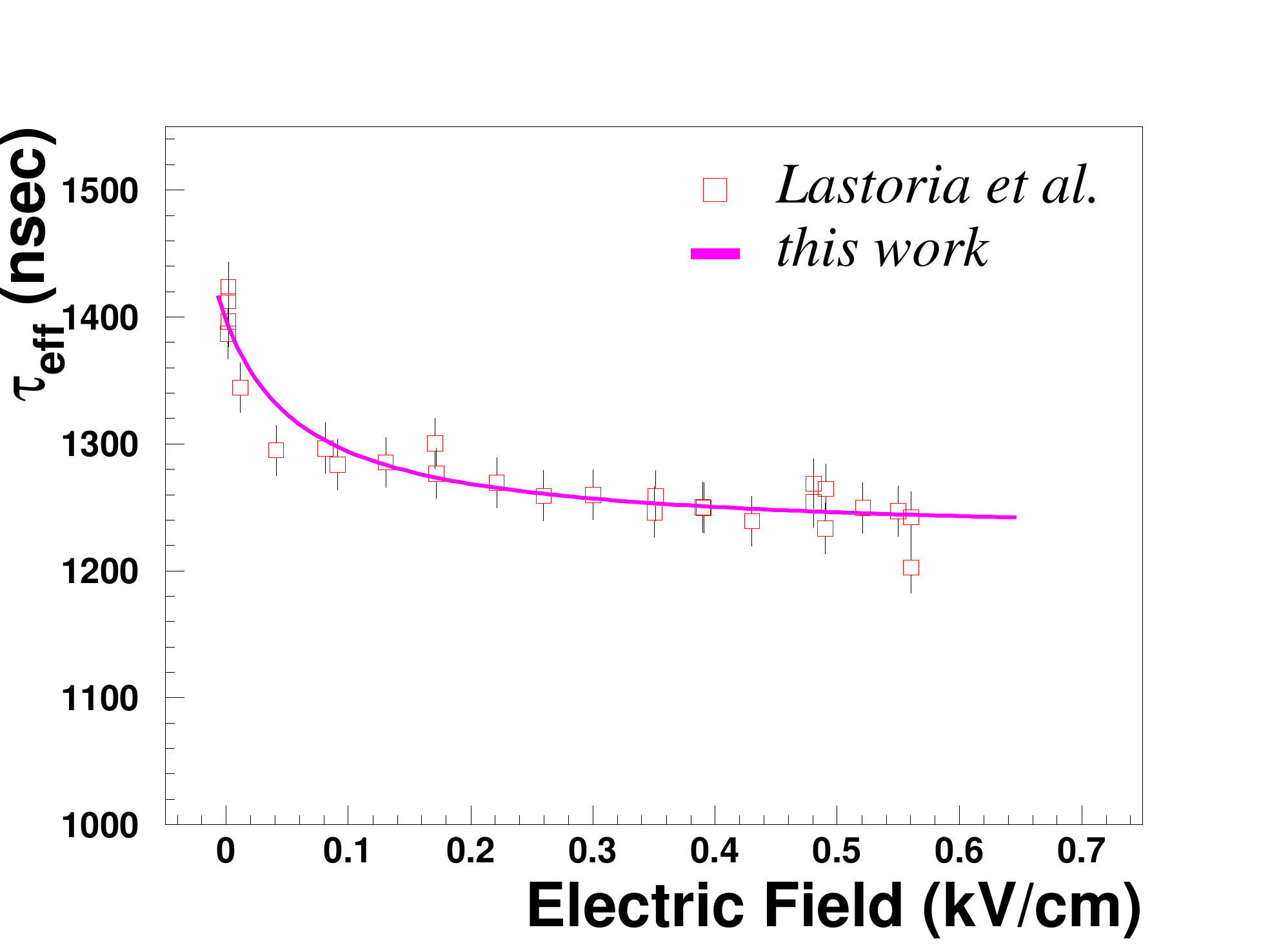}
	\caption{\label{fig:tau-ef} Variation of the decay time of the slow scintillation component, $\tau_{eff}$ as a function of the applied electric field. Magenta line represents a fit of the data with the function of equation \ref{eq:tau_eff}.}  
\end{figure}

The parameters k$^+$ and q can be written as:

\begin{eqnarray}
\label{eq:k+}
k^+(\mathcal{E}) =  k^+_0\Big[1+\frac{A}{B(1+\nicefrac{k_\mathcal{E}}{\mathcal{E}})}\Big]\\
q(\mathcal{E}) = q_0\Big[1-\frac{A}{(A+\nicefrac{N_{ex}}{N_i})(1+\nicefrac{k_\mathcal{E}}{\mathcal{E}})}\Big]
\end{eqnarray} 

where k$_0^+$ and q$_0$ are the values of k$^+$ and q at zero electric field and at a given value of LET. For low LET particles, for which the phenomenon of escaping electrons is present ($k^+\neq 0$) and when $q~\tau_q \ll 1$, the scintillation signal of equation \ref{eq:l_t} presents only a small deviation from a purely exponential decay with a characteristic time of $\tau_{eff}\simeq\tau_q$ and its dependence on the electric field can be explicitly written as:
\begin{equation}
\label{eq:tau_eff}
\tau_{eff} \simeq \tau_q = \frac{1}{\frac{1}{\tau_3}+k_0^++\frac{k_0^+ A}{B(1+\nicefrac{k_\mathcal{E}}{\mathcal{E}})}}=\frac{1}{\alpha+\frac{\beta}{1+\nicefrac{k_\mathcal{E}}{\mathcal{E}}}}
\end{equation}

where $\alpha$=$\nicefrac{1}{\tau_{eff}(0)}$ is the inverse of the characteristic time at zero electric filed and $\beta=\nicefrac{k_0^+ A}{B}$.\\
The measurement of the variation of the slow decay time of the LAr scintillation light as a function of the applied electric field, reported in \cite{Lastoria:2019qiw}, represents, substantially, a measurement of the electron-ion recombination process in LAr performed with light. Experimental points taken from \cite{Lastoria:2019qiw} are shown in figure \ref{fig:tau-ef}. 
They have been fitted with the function of equation \ref{eq:tau_eff}, leaving the parameters $\alpha$, $\beta$ and k$_E$ free and the result is shown with a magenta line. The agreement between data and the model is pretty good, with the exception of one of the points at very low electric field. The fit procedure returns a value of k$_\mathcal{E}$~=~0,075~$\pm$~0,015~kV/cm, which is compatible with the value of 0,0486~$\pm$~0,0006~kV/cm~$\frac{\nicefrac{g}{cm^2}}{MeV}$ reported in \cite{AMORUSO2004275}, when considering cosmic muons with energies between 1~GeV and 4~GeV and a stopping power between 1,6~$\frac{MeV}{\nicefrac{g}{cm^2}}$ and 2,0~$\frac{MeV}{\nicefrac{g}{cm^2}}$ in LAr.

\section{Xenon doping of LAr}
It is known that adding xenon to LAr at the level of ppm (part per million) has the effect of shifting the wavelength of the slow scintillation component from 127~nm to 174~nm \cite{Wahl_2014}. The complete shift of the slow component is observed at tens of ppm of xenon concentration. It has been also reported an increase of the number of detected photons with increasing xenon concentration, which can not be explained with an increase of the conversion efficiency of the TPB, the wavelength shifter used in the experiment, which has been measured to be almost the same at 127~nm and 174~nm \cite{TPB-benson}.\\     
The mechanism suggested in \cite{KUBOTA199371} for the transfer of the excitation energy from argon to xenon can be summarized with the following reaction chain:
\begin{eqnarray}
\label{eq:ar_2toarxe}
Ar_2^* + Xe \rightarrow (ArXe)^* + Ar\\
\label{eq:arxetoxe2}
(ArXe)^* + Xe\rightarrow  Xe_2^* + Ar
\end{eqnarray} 

The energy transfer process and the subsequent emission of 174 nm photons compete with the radiative decay of the Ar$_2^*$ and with the quenching processes described in section \ref{model}. The net effect is the shift of the slow LAr scintillation component and the partial recovery of the quenched LAr species, which both result in the emission of 174~nm photons with peculiar time characteristics.\\
In order to understand the gross features of the light emission process from a xenon-argon mixture in the case of low LET particles,  it is worth making some rough approximations. 
For a high enough Xe concentration, it should be possible to neglect second order quenching effects, such as Ar$_2^*$-Ar$_2^*$, Ar$_2^*$-(ArXe)$^*$, (ArXe)$^*$-(ArXe)$^*$. Assuming that the energy transfer between argon and xenon happens without losses and that the reaction rates of the processes Ar$_2^*$ + Xe $\rightarrow$ (ArXe)$^*$ and (ArXe)$^*$ + Xe $\rightarrow$ Xe$_2^*$ are the same and equal to k$_{Xe}$,
the instantaneous variation of the density of triplet argon states, N$_3^{Ar}$(t), of Xe$_2^*$ dimers, N$^{Xe}$(t) \footnote{Xe$_2^*$ singlet and triplet states are treated together here, since their decay times are very close \cite{DOKE199962}.} and of mixed dimers (ArXe)$^*$, N$^{ArXe}$(t)  can be written as:
\begin{widetext}
\begin{eqnarray}
\label{eq:dnar}
\frac{dN_3^{Ar}(t)}{dt} = -\lambda_3~N_3^{Ar}(t)-k^+N_3^{Ar}(t)-k_{Xe} [Xe]N_3^{Ar}(t)\\
\label{eq:dnarxe}
\frac{dN^{ArXe}(t)}{dt} = -k_{Xe} [Xe] N^{ArXe}(t) + k_{Xe} [Xe] N_3^{Ar}(t) \\
\label{eq:dnxe}
\frac{N^{Xe}(t)}{dt} = -\lambda_{Xe} N^{Xe}(t) +  k_{Xe} [Xe] N^{ArXe}(t)
\end{eqnarray}
\end{widetext}

where $\lambda_{Xe}$ is the inverse of the characteristic time of xenon emission and [Xe] is the xenon concentration.
Assuming that $\lambda_{Xe}$ $\gg$ k$_{Xe}$[Xe], equations \ref{eq:dnar}, \ref{eq:dnarxe} and \ref{eq:dnxe} can be easily solved with the initial condition that N$^{ArXe}$(0)~=~0 and give the probability density for the slow LAr scintillation component, l$_3$(t), and for Xe shifted light, l$_{Xe}(t)$:
\begin{eqnarray}
\label{eq:l_3_t}
l_3(t) = \frac{\alpha_3}{\tau_3}~e^{-\frac{t}{\tau_r}}\\
\label{eq:l_xe_t}
l_{Xe}(t) = \alpha_3 (k_{Xe}[Xe])^2 \tau_q [e^{-\frac{t}{\tau_d}}-e^{-\frac{t}{\tau_r}}]
\end{eqnarray}        

where $\tau_r = 1/(k_{Xe}[Xe]+\nicefrac{1}{\tau_q})$ and $\tau_d = 1/k_{Xe}[Xe]$. The probability density for all the emitted photons, regardless of their wavelength, can be obtained by summing the contributions of equations \ref{eq:l_3_t} and \ref{eq:l_xe_t} to the fast LAr scintillation component, assumed to be unaffected by xenon doping:
\begin{equation}
\label{eq:l_tot_t}
l(t) = \frac{\alpha_s}{\tau_s}e^{-\frac{t}{\tau_s}} + l_3(t) + l_{Xe}(t)
\end{equation}
The total amount of emitted light is obtained by integrating equation
\ref{eq:l_tot_t}:
\begin{equation}
L = \alpha_s + \alpha_3 \frac{\tau_r}{\tau_3} + \alpha_3 \frac{(k_{Xe}[Xe])}{(k_{Xe}[Xe])+\nicefrac{1}{\tau_q}}
\end{equation}

Using the parameters of the scintillation waveform for electron recoils found in section \ref{sec:electron_neutrons}, and the reaction rate k$_{Xe}$~=~8,8$\times$10$^{-5}$~ppm$^{-1}$nsec$^{-1}$ (with ppm in mass)
reported in \cite{Wahl_2014} it is possible to predict approximately the shape of the scintillation signal for electron recoils and the dependence of the LY on the xenon concentration, [Xe].\\
\begin{figure}
	\includegraphics[width=10cm]{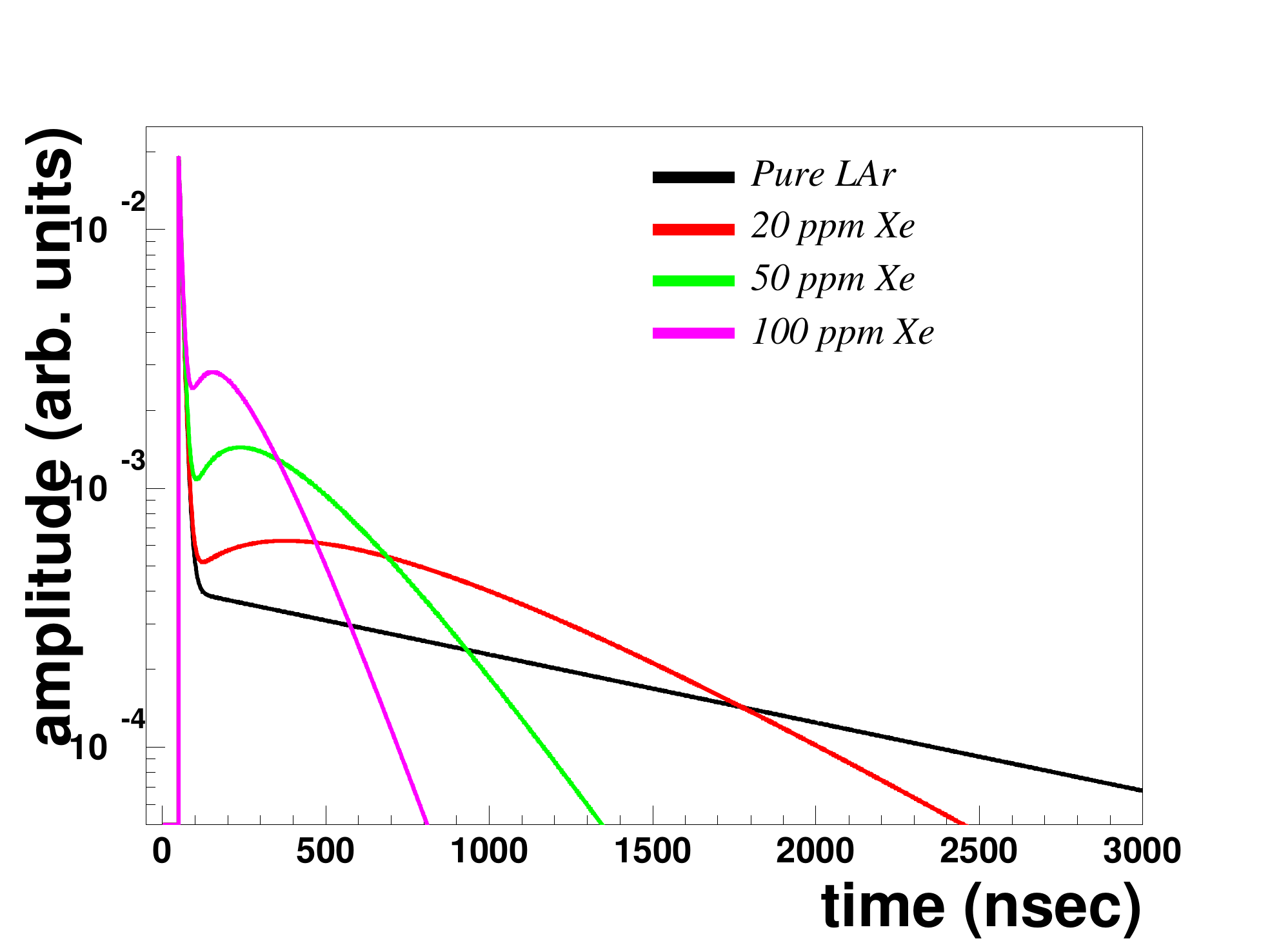}
	\caption{\label{fig:xe-wave} Waveforms of xenon doped liquid argon at different xenon concentrations as predicted by the model. The waveforms represent the sum of LAr and xenon shifted light.}  
\end{figure}
Few waveforms for different xenon concentrations are shown in figure \ref{fig:xe-wave}. The exact shape will depend on the precise values of the reaction rates and on the eventual conversion of fast (and slow) LAr scintillation light through photo-absorption by xenon atoms. This would lead to the formation of (ArXe)$^*$  which would evolve in Xe$_2^*$ according to reaction \ref{eq:arxetoxe2}. 
   
\begin{figure}
	\includegraphics[width=10cm]{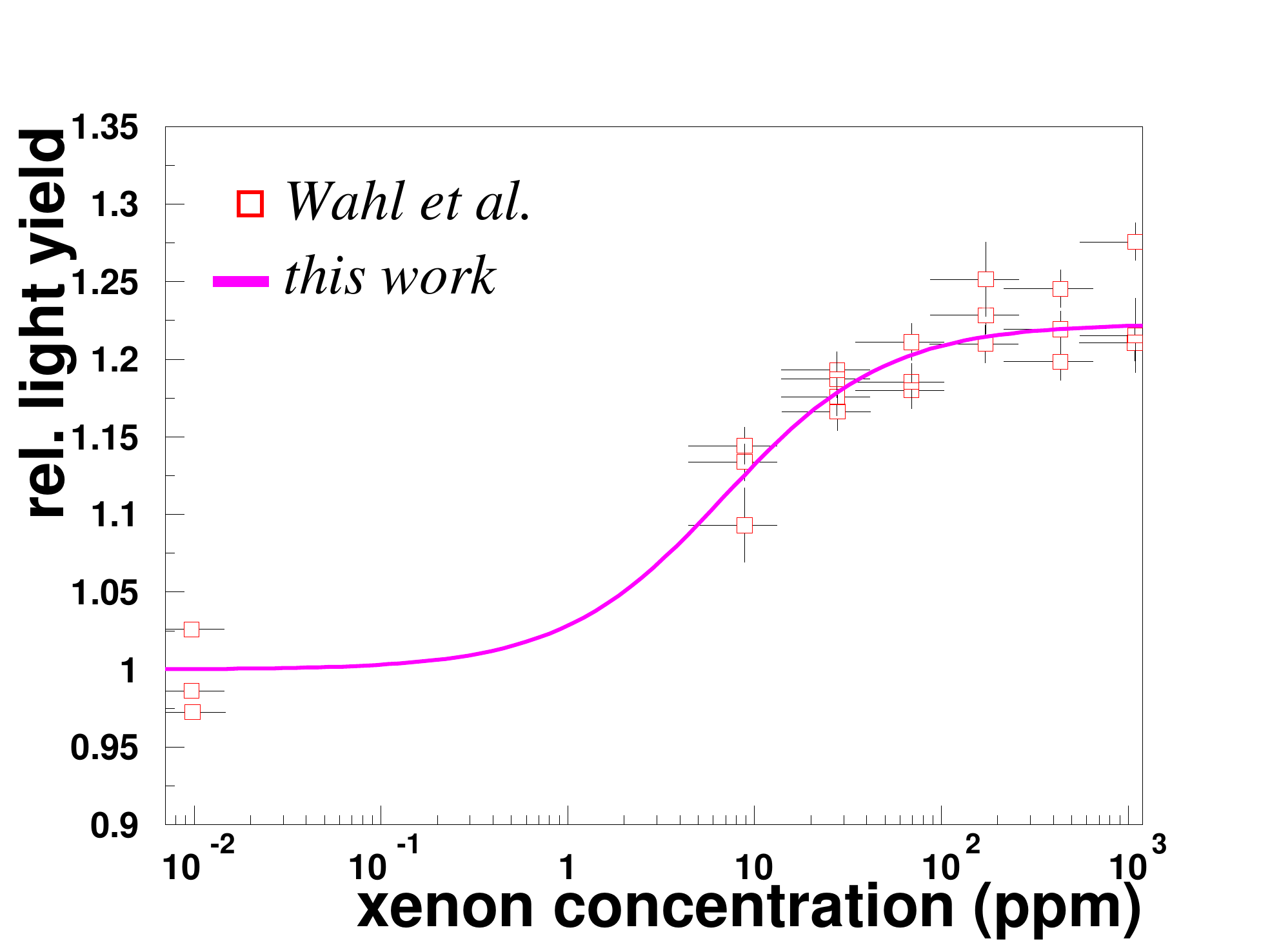}
	\caption{\label{fig:xe-ly} Variation of the LY of LAr and xenon shifted photons as a function of the xenon concentration in pppm (mass). The experimental points at zero xenon concentration have been shifted to 10$^{-2}$~ppm to facilitate the visualization. The model prediction is shown as a magenta line.}  
\end{figure}

The variation of the total LY, which includes 127~nm and 174~nm photons, predicted by the model is shown in figure \ref{fig:xe-ly} together with the experimental data shown in \cite{Wahl_2014}. Both model prediction and data have been normalized by their value at zero xenon concentration. An increase of the overall LY around 25\% for concentrations above few tens of ppm in mass is observed in the data and correctly predicted by the model. It is reasonable to conclude that the two main approximations made: neglecting second order quenching effects and assuming a lossless transfer of energy between xenon and argon are small and compensate with each other, leading to a good description of the experimental data. The first approximation would lead to an increase of LY, since the amount of quenched Ar$_2^*$ would be higher, while the second would lead to a decrease of LY, because of possible non-radiative dis-excitations of the (ArXe)$^*$ states.  

\section{F-PROMPT for electron and nuclear recoils}
\label{sec:fprompt}
LAr allows for a powerful particle discrimination based on the shape of its scintillation signal. In particular, the relative abundances of fast and slow components strongly depend on the particle type \cite{Doke:1988rq}. This property of LAr is crucial for discarding gamma and electron backgrounds from nuclear recoil events in direct Dark Matter experiments (\cite{ACCIARRI20121113}, \cite{WArP}, \cite{PhysRevD.93.081101-DS50}, \cite{ArDM}, \cite{PhysRevC.78.035801-lippi}).\\
The variable which is typically used to discriminate electron from nuclear recoils in LAr is F$_{prompt}$, which informs the abundance of the fast component in the scintillation signal and is defined as:
\begin{equation}
\label{eq:fprompt}
F_{prompt}=\frac{\int_{0}^{t^*}l(t)~dt}{\int_{0}^{\infty}l(t)~dt}
\end{equation}
where l(t) is the scintillation waveform and t$^*$ is the integration time of the fast component which maximizes the separation. Different values of t$^*$ are used by different groups, but it is typically close to 100~nsec. 
F$_{prompt}$ depends on the recoil energy and the separation between electron and nuclear recoils tends to get worse at lower energies. The proposed model contains explicitly the dependence of the scintillation signals of electron and nuclear recoils from the density of deposited energy and can be used to predict and explain the behavior of F$_{prompt}$ observed experimentally.\\

In order to evaluate F$_{prompt}$ for electrons as a function of the kinetic energy, E, of the incoming electron, F$_{prompt}^e(E)$, it is necessary to evaluate the parameters k$^+$ and q of equation \ref{eq:l_t}. The prameter k$^+$ is proportional to the number density N$_0^+$ of Ar$_2^+$, whose dependence on the density of deposited energy is not known. It is reasonable to assume, in first approximation, that it stays constant in the region of interest, below 1 MeV. The density of deposited energy varies very slowly down to 100~keV, where it starts increasing more steeply. In this low energy region the effect of the increased density is compensated by the reduction of the escaping probability for the ionization electrons due to the increased electric field in the core of the track.\\
The parameter q is equal to:
\begin{equation}
\label{eq:q_electrons}
q^e = \sigma_3 v_3 N_0 = \sigma_3 v_3 \Big[\frac{dE}{dx}\frac{1}{r_3^2}\Big] \frac{\alpha_3^e}{W_{el}}\Big(1-R(0)+\frac{N_{ex}}{N_i}\Big)
\end{equation}
where  $\nicefrac{dE}{dx}$ is the electronic stopping power of argon, $[\nicefrac{dE}{dx~r_3^2}]$ is the average value of the density of deposited energy along the track, W$_{el}$~=~23,6~eV is the average energy to produce an electron ion pair in LAr \cite{Doke:1988rq}, R(0) is the recombination factor at zero electric field and $\alpha_3^e$ = 0,85 is the initial relative abundance of triplet states for electron recoils. For low energy electrons ($<$~1~MeV),  according to Bohr's theory \cite{Leo:1987kd}, 
$r_3$ is proportional to $\gamma~\beta$, in particular:
\begin{equation}
\label{eq:energy_density_electrons}
\frac{dE}{dx} \frac{1}{r_3^2} \simeq \frac{dE}{dx} \frac{\bar{\nu}^2}{c^2 \beta^2 \gamma^2}
\end{equation} 
where c is the speed of light, $\beta$=v/c, $\gamma$ = 1/$\sqrt{(1-\beta^2)}$ and $\bar{\nu}$ is the average orbiting frequency of atomic electrons, which is related to the mean excitation potential, I, by the relation $I=h~\bar{\nu}$. For argon, I = 188 eV.
Using equations \ref{eq:q_electrons} and \ref{eq:energy_density_electrons}, $q$ is written as:
\begin{equation}
q^e(E)= k_q^ed^e(E)
\end{equation}
where $d^e(E)$ is the average value along the track of the density of deposited energy of equation \ref{eq:energy_density_electrons} multiplied by $\alpha_3^e$:
\begin{equation}
\label{eq:energy_density_electron_average}
d^e(E) = \int_{0}^{E} \frac{dE'}{dx} \frac{\alpha_3^e~\bar{\nu}^2}{c^2\beta^2 \gamma^2} \frac{dE'}{E}
\end{equation}
and k$_q^e$ is a characteristic constant:
\begin{equation}
\label{eq:k_q^e}
k_q^e = \frac{\sigma_3 v_3}{W_{el}}\Big(1-R(0)+\frac{N_{ex}}{N_i}\Big)
\end{equation}
In order to compare the model to data, the integral of equation \ref{eq:energy_density_electron_average} is evaluated numerically using tabulated values for the electronic stopping power \cite{ESTAR}.

\begin{figure}
	\includegraphics[width=10cm]{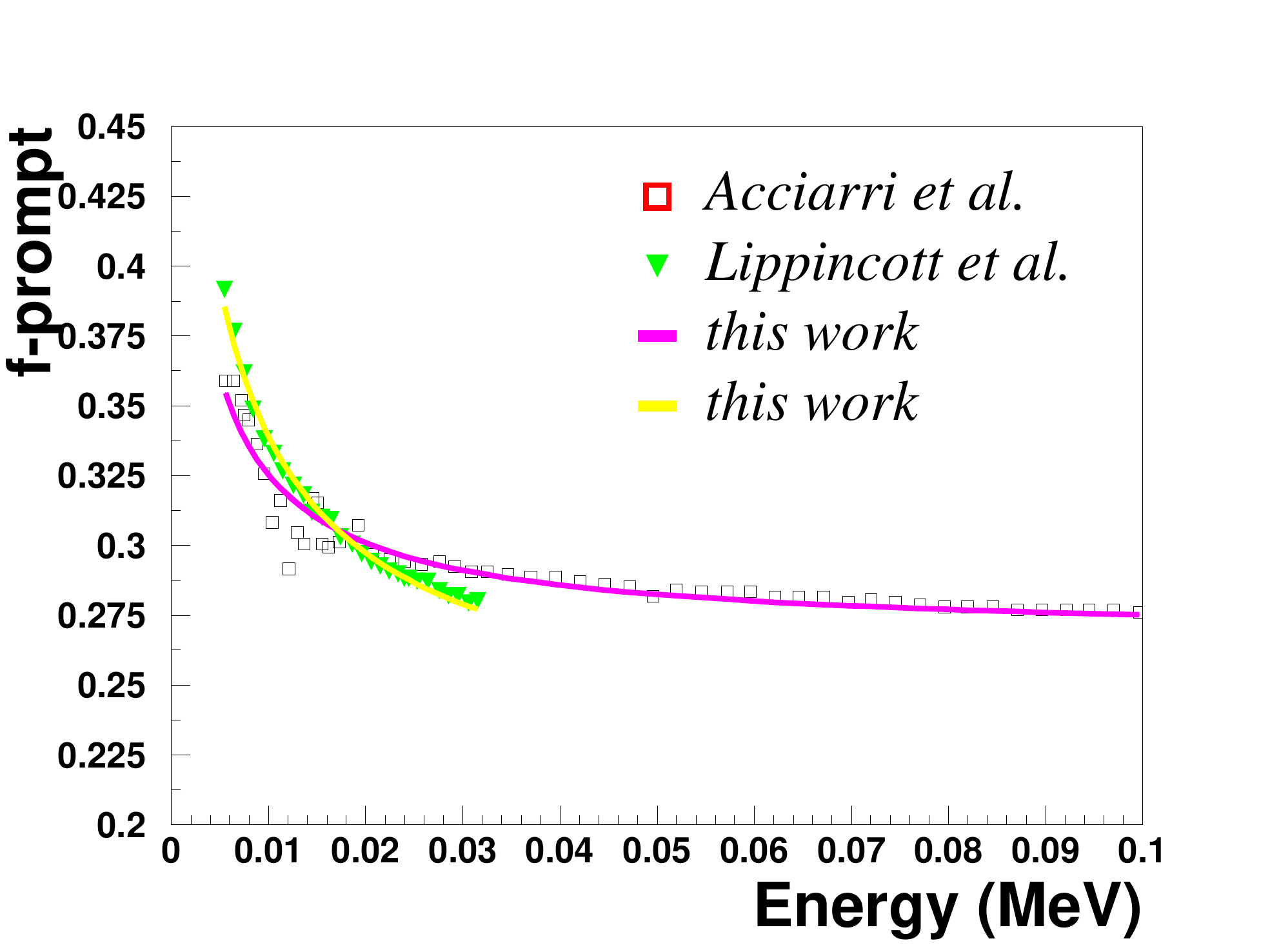}
	\caption{\label{fig:fprompt_el}F$_{prompt}$ for low energy electrons in LAr measured by two experimental groups \cite{ACCIARRI20121113} \cite{PhysRevC.78.035801-lippi}. The experimental points have been fitted with equation \ref{eq:f_prompt_calculated} with fitting parameters p$_1$, p$_2$ and k$_q$ (see text). The results of the fit procedures are shown with magenta (\cite{ACCIARRI20121113}) and green (\cite{PhysRevC.78.035801-lippi}) lines.}  
\end{figure}

$d^e(E)$ is well fitted, for E$<$1~MeV, by the analytical expression:
\begin{equation}
d^e(E)= 843 (E^{-0.65}-1.48 E^{-0.43}+0.66)
\label{eq:de_numerical}
\end{equation}
with E in MeV and d$^e$ in MeV/$\mu$m$^{3}$.
The model prediction is compared to experimental data from \cite{ACCIARRI20121113} and \cite{PhysRevC.78.035801-lippi}. F$_{prompt}^e$ is calculated as:
\begin{equation}
\label{eq:f_prompt_calculated}
F_{prompt}^{e,n}= \frac{p_1 L_s + p_2 L3}{L_s + L_3}
\end{equation} 
where L$_1$ and L$_3$ are the abundances of the fast and slow components respectively (see equation \ref{eq:light_partition}), the parameters p$_1$ and p$_2$ are related to the integration process for the calculation of F$_{prompt}^{e,n}$ (see equation \ref{eq:fprompt}) and in particular to the fraction of fast and slow component which fall inside the integration window. Typically p$_1$ is close to one and p$_2$ is close to zero. The fact of p$_1$ not being exactly equal to one is attributed to the delayed light emission of the wavelength shifters used to detect LAr photons \cite{PhysRevC.91.035503}. The parameters p$_1$, p$_2$ and k$_q^e$ are left free and adjusted on data. The two data sets have been fitted separately, since they show some differences in their asymptotic behaviors at large energies, which is probably related to  different integration intervals, and in their slopes at low energies. 
The result of the fitting procedures is shown in figure \ref{fig:fprompt_el}. The values of $k_q^e = 0.73\times 10^{-7}$~nsec$^{-1}$~MeV$^{-1}$~${\mu}m^3$ for \cite{ACCIARRI20121113} and $k_q^e = 1.9\times 10^{-7}$ nsec$^{-1}$~MeV$^{-1}$~${\mu}m^3$ for \cite{PhysRevC.78.035801-lippi} are obtained.\\

In the case of a low energy  nuclear recoils of an argon atom in LAr, below few hundreds of keV, a significant amount of the energy lost by the recoiling argon atom is due to elastic collisions with other argon nuclei and only a small fraction of it is transferred to the electrons. The Lindhard theory \cite{Lindhard} \cite{Hitachi2019PropertiesFL} gives the amount of energy transferred to the electrons of the LAr in terms of the dimensionless variable $\varepsilon$:
\begin{equation}
\label{eq:epsilon}
\varepsilon = C_{\varepsilon}E = \frac{a_{TFF}A_2}{Z_1 Z_2 e^2 (A_1+A_2)}E
\end{equation}
where E is the recoil energy, Z and A are the atomic and mass number of the projectile (1) and of the medium (2) and:
\begin{equation}
a_{TFF}=\frac{0.8853~a_B}{(Z_1^{1/2}+Z_2^{1/2})^{2/3}}
\end{equation} 
a$_B=\hbar/m_e e^2 = 0.529~\AA$ is the Bohr radius. For Z$_1$ = Z$_2$ equation \ref{eq:epsilon} gives C$_{\varepsilon}$ = 0.01354~keV$^{-1}$.
The amount of energy transferred to the electrons is given by \cite{Mei2008AMO}:
\begin{equation}
\label{eq:eta_epsilon}
\eta(\varepsilon)=\frac{k~g(\varepsilon)~\varepsilon}{1+k~g(\varepsilon)}
\end{equation}
where k = 0.133~Z$^{2/3}$~A$^{-1/2}$ = 0.144 and g($\varepsilon$) is fitted with the function \cite{Mei2008AMO}:
\begin{equation}
g(\varepsilon)=3 \varepsilon^{0.15} + 0.7 \varepsilon^{0.6} +\varepsilon 
\end{equation}
Low energy nuclear recoils deposit their energy in a very confined portion of space. The projected range for argon ions of tens of keV kinetic energy in LAr is of the order of few hundreds of $\AA$ \cite{ZIEGLER20101818} and the transversal dimensions of the core of the track is of the order of the Bohr radius a$_B$ \cite{Hitachi2019PropertiesFL}. It is reasonable to assume that in this range of energies, the spatial distribution of the Ar$_2^*$ is largely dominated by the fast diffusion of excitons and holes before self-trapping and by the coulombian repulsion of the Ar ions in the core of the track. Under this hypothesis, the recoil energy is deposited inside an approximately constant volume at very low energies. A linear growth of this volume with energy is expected, since the total stopping power is constant for energies below few hundreds ok keV \cite{Lindhard} \cite{Hitachi2019PropertiesFL} and the range of the Ar ions is proportional to its initial kinetic energy, while the transversal dimensions continue being dominated by fast diffusive processes. Using equations \ref{eq:q_electrons} and \ref{eq:eta_epsilon}, the factor {\it q} for low energy nuclear recoils is written as:
\begin{equation}
\label{eq:q_n_E}
q^n(E) = k_q^n~d^n(E) = k_q^n\frac{\alpha_3^n}{V_0}\frac{\eta(E)}{1+k_V E}
\end{equation}

where $\alpha_3^n$ = 0.35 is the initial relative abundance of triplet states for nuclear recoils, V$_0$ is the volume inside which the energy deposited by nuclear recoils is contained for very low energies, k$_V$ takes into account the increase of the volume with the recoil energy and k$_q^n$ is given by: 
\begin{equation}
\label{eq:k_q^n}
k_q^n = \frac{\sigma_3 v_3}{W_{el}}\Big(1+\frac{N_{ex}}{N_i}\Big)
\end{equation}
Equations \ref{eq:q_n_E} and \ref{eq:light_partition} (with $\tau_q$ = $\tau_3$) can be substituted in equation \ref{eq:f_prompt_calculated} to evaluate F$_{prompt}^n$ for nuclear recoils and compare it to available data.\\
The model is fitted to the data reported in \cite{ACCIARRI20121113} and referred to, as the {\it high light yield} sample, which span a broad interval of energies, from 10~keV to above 1~MeV.\\
The conversion between the number of detected photo-electrons (N$_{phel}$) and the nuclear recoil energy in \cite{ACCIARRI20121113} is done assuming a constant relative scintillation yield between nuclear and electron recoils, $\mathcal{L}_{eff}$, equal to 0,3. A more appropriate conversion between N$_{phel}$ and the recoil energy, E$_n$, is given by:
\begin{equation}
\label{eq:nphel_energy}
E_n = \frac{N_{phel}}{LY\times\mathcal{L}_{eff}(E_n)}
\end{equation}
where LY is the Light Yield for electron recoils expressed in phel/keV. The fit procedure of the F$_{prompt}^n$ data reported in \cite{ACCIARRI20121113}, which takes into account equation \ref{eq:nphel_energy} is described in the following section.  

\begin{figure}
	\includegraphics[width=10cm]{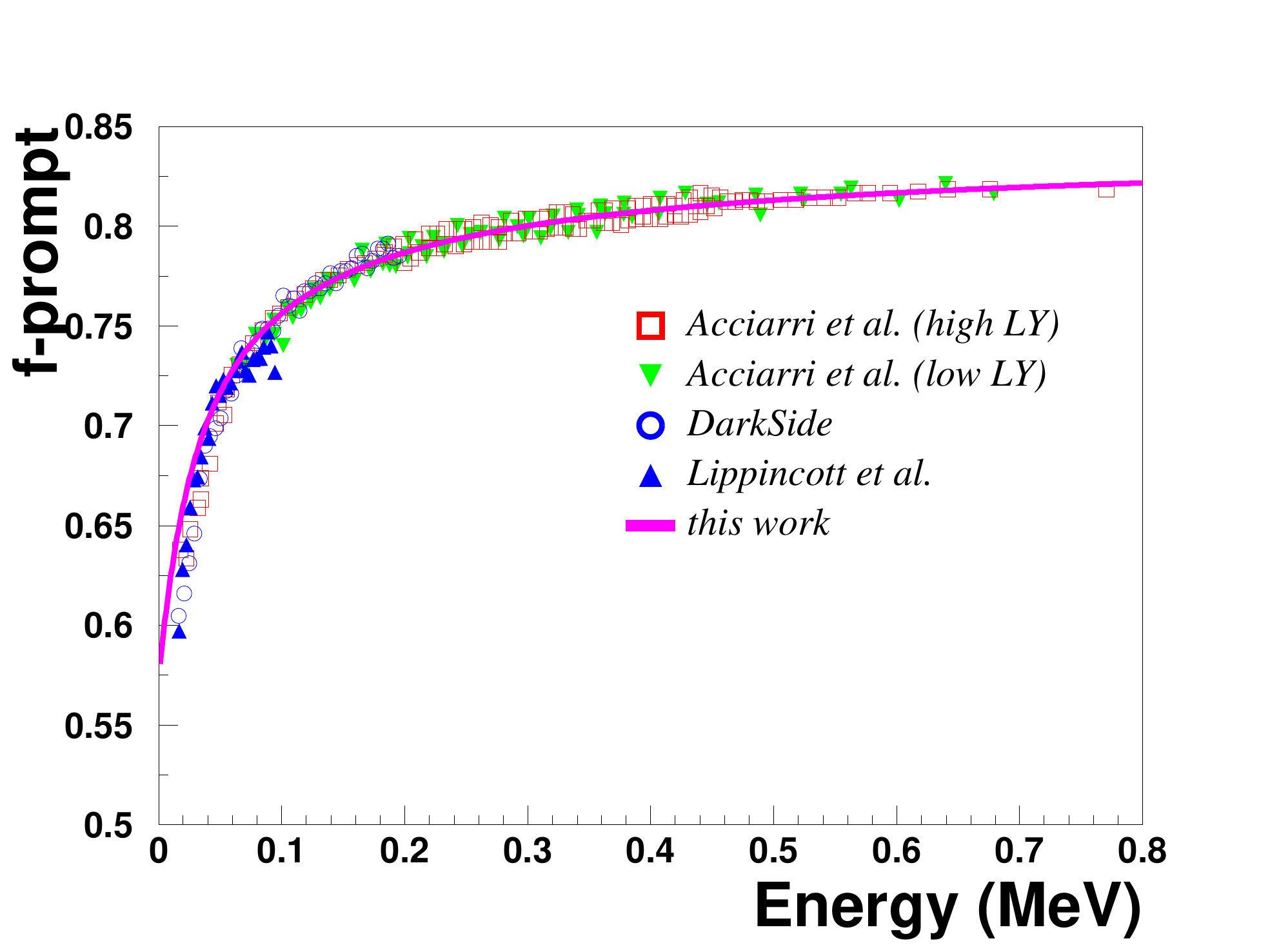}
	\caption{\label{fig:fprompt_n}F$_{prompt}$ for nuclear recoils in LAr measured by different authors \cite{ACCIARRI20121113} \cite{PhysRevC.78.035801-lippi} \cite{PhysRevD.93.081101-DS50}. The high light yield sample (see text) from \cite{ACCIARRI20121113} is used to fit the data. Magenta line represents the fit result. The data sets have been aligned by applying overall scale factors which vary between 1.01 and 1.05.}  
\end{figure}

\section{Quenching factor for nuclear recoils}
\label{sec:quenching-nuclear-recoils}
The quenching of the number of emitted photons in a nuclear recoil event is described as the succession of three distinct quenching processes: (i) the quenching of the amount of energy transferred to the atomic electrons, due to the elastic collisions of the argon ion with surrounding argon atoms \cite{Lindhard}; (ii) the quenching of the excitons formed after the nuclear recoil due to bi-excitonic quenching \cite{Hitachi2019PropertiesFL}; (iii) the quenching of the triplet states formed after the trapping of the excitons due to triplet-triplet interactions (and triplet-ion in the presence of Ar$_2^+$). The overall quenching factor for nuclear recoils can be written as:
\begin{equation}
Q^N = Q^L\times Q^E\times Q^T
\end{equation} 
Q$_L$ is the quenching factor of process (i), which has been calculated by Lindhard \cite{Lindhard} and can be written as (refer to equation \ref{eq:eta_epsilon}):
\begin{equation}
\label{eq:Q_L_Lindhard}
Q^L = \frac{\eta(\varepsilon)}{\varepsilon} = \frac{k~g(\varepsilon)}{1+k~g(\varepsilon)}
\end{equation}
Q$^E$ is the quenching factor of process (ii). It has been pointed out  in \cite{PhysRevB.46.11463} that Q$^E$ can be considered  approximately constant for energies below few hundreds of keV. This approximation has been proven to work well for liquid xenon \cite{HITACHI2005247-xenon}. 
Q$^T$ is the quenching factor for process (iii), discussed in this work. It can be written as (see equation \ref{eq:light_partition}):
\begin{equation}
\label{eq:Q_T_segreto}
Q^T= \alpha_s + \alpha_3 \frac{ln(1+q^n(E)~\tau_3)}{q^n(E)~\tau_3}
\end{equation}
where $\alpha_s$~=~0.65, $\alpha_3$~=~0.35, $\tau_3$~=~2100~nsec and q$^n$(E) is given by equation \ref{eq:q_n_E}.\\

Experimentally, it is convenient to measure the relative scintillation yield of nuclear recoils with respect to electron recoils, $\mathcal{L}_{eff}$. Nuclear recoils are typically compared to the recoils of electrons emitted after the photo-absorption of the $\gamma$ lines of $^{241}$Am (59.5~keV) and $^{57}$Co (122~keV). $\mathcal{L}_{eff}$ can be estimated as:
\begin{equation}
\label{eq:leff}
\mathcal{L}_{eff} = \frac{Q^N}{Q^{el}} = Q^L\times Q^T \times \frac{Q^E}{Q^{el}}
\end{equation}
where Q$^{el}$ is the quenching factor of the reference electron recoil. Since Q$^{E}$ and Q$^{el}$ are both constant, the ratio $\nicefrac{Q^E}{Q^{el}}$ is an overall multiplicative constant in equation \ref{eq:leff}.\\
In order to estimate the parameters of the model, an overall fit of the F$_{prompt}^n$ data from \cite{ACCIARRI20121113} and of the $\mathcal{L}_{eff}$ data reported by the ARIS Collaboration in \cite{PhysRevD.97.112005} is performed, where the product $\nicefrac{k_q \alpha_3^n}{V_0}$ and k$_V$ from equation \ref{eq:q_n_E}  and the ratio $\nicefrac{Q^E}{Q^{el}}$ of equation \ref{eq:leff}  are left as free parameters. The exact value of E$_n$ for the F$_{prompt}^n$ set of data is obtained by inverting numerically equation \ref{eq:nphel_energy}. 
The fit procedure returns a value of 0,11$\pm$0,01~MeV$^{-1}$~nsec$^{-1}$ for the product $\nicefrac{k_q \alpha_3^n}{V_0}$, 9,1$\pm$0,1~Mev$^{-1}$ for k$_V$  and 1,00$\pm$0,01 for the ratio $\nicefrac{Q^E}{Q^{el}}$. Taking into account the value of k$_q^e$ found with the fit of F$_{propmpt}^e$ for electron recoils, that $\alpha_3^n$~=0.35 and that the recombination factor for nuclear recoils at zero electric field is zero (R(0)=0),  the volume V$_0$ ranges between 200 and 500 nm$^3$. This corresponds to a sphere with a diameter between 7 and 10~nm, which is in a reasonable agreement with the diffusion of free excitons and holes before trapping, for times of the order of 1~psec and assuming a diffusion constant D~=~1~cm$^2$/sec \cite{Hitachi2019PropertiesFL}.\\
   
A fit procedure which uses only the F$_{prompt}^n$ data set is able to constrain pretty well the terms $\nicefrac{k_q \alpha_3^n}{V_0}$ and k$_V$, giving results compatible with the ones reported here, while the range for the ratio $\nicefrac{Q^E}{Q^{el}}$ turns out to be quite broad (between 0,9 and 1,1). The same ratio $\nicefrac{Q^E}{Q^{el}}$ could be estimated following the $\alpha$-core approximation discussed in \cite{Hitachi2019PropertiesFL} giving a value close to unity, but with a quite large uncertainty (at the level of 20-30\%).\\        

The result of the fit for F$_{prompt}^n$ is shown in figure \ref{fig:fprompt_n} together with three more data sets. The data set referred as {\it low light yield} is also taken from \cite{ACCIARRI20121113}, but it is obtained with a different experimental set-up, with a lower light yield with respect to the {\it high light yield} one. The two other data sets are taken from \cite{PhysRevD.93.081101-DS50} and \cite{PhysRevC.78.035801-lippi}

The data sets have been aligned by applying overall scaling factors to the F$_{prompt}^n$ values (one per each data set), which range between 1.01 and 1.05. The not perfect overlap of the different data sets before rescaling is attributed to different integration intervals to compute F$_{prompt}^n$ and to small systematic effects.\\

\begin{figure}
	\includegraphics[width=10cm]{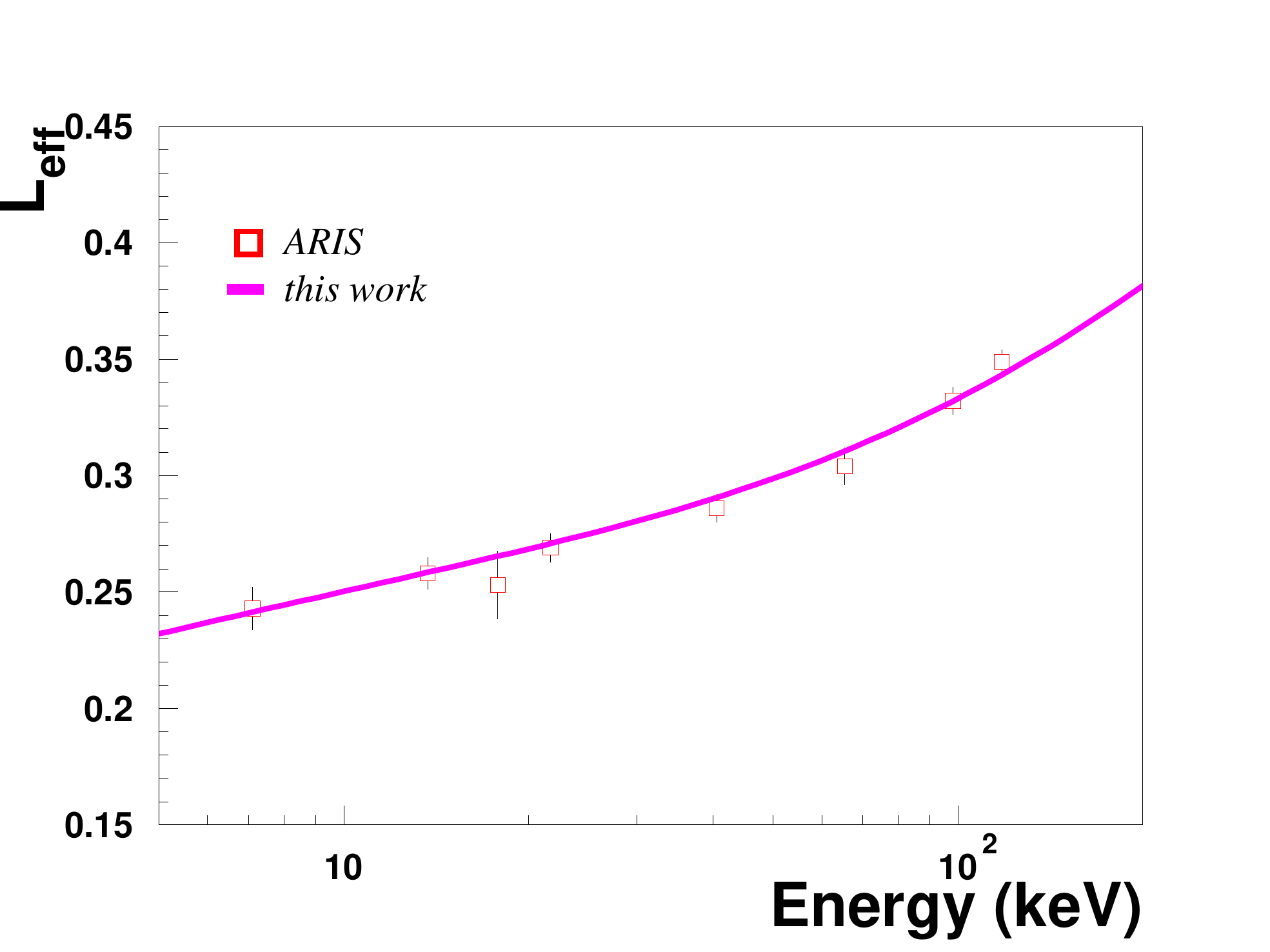}
	\caption{\label{fig:leff_aris} Comparison of the model prediction for $\mathcal{L}_{eff}$ (magenta line) with the data from the ARIS Collaboration \cite{PhysRevD.97.112005}. The factor $\nicefrac{Q^E}{Q^{el}}$ has been adjusted to data (see text).}  
\end{figure}

\begin{figure}
	\includegraphics[width=10cm]{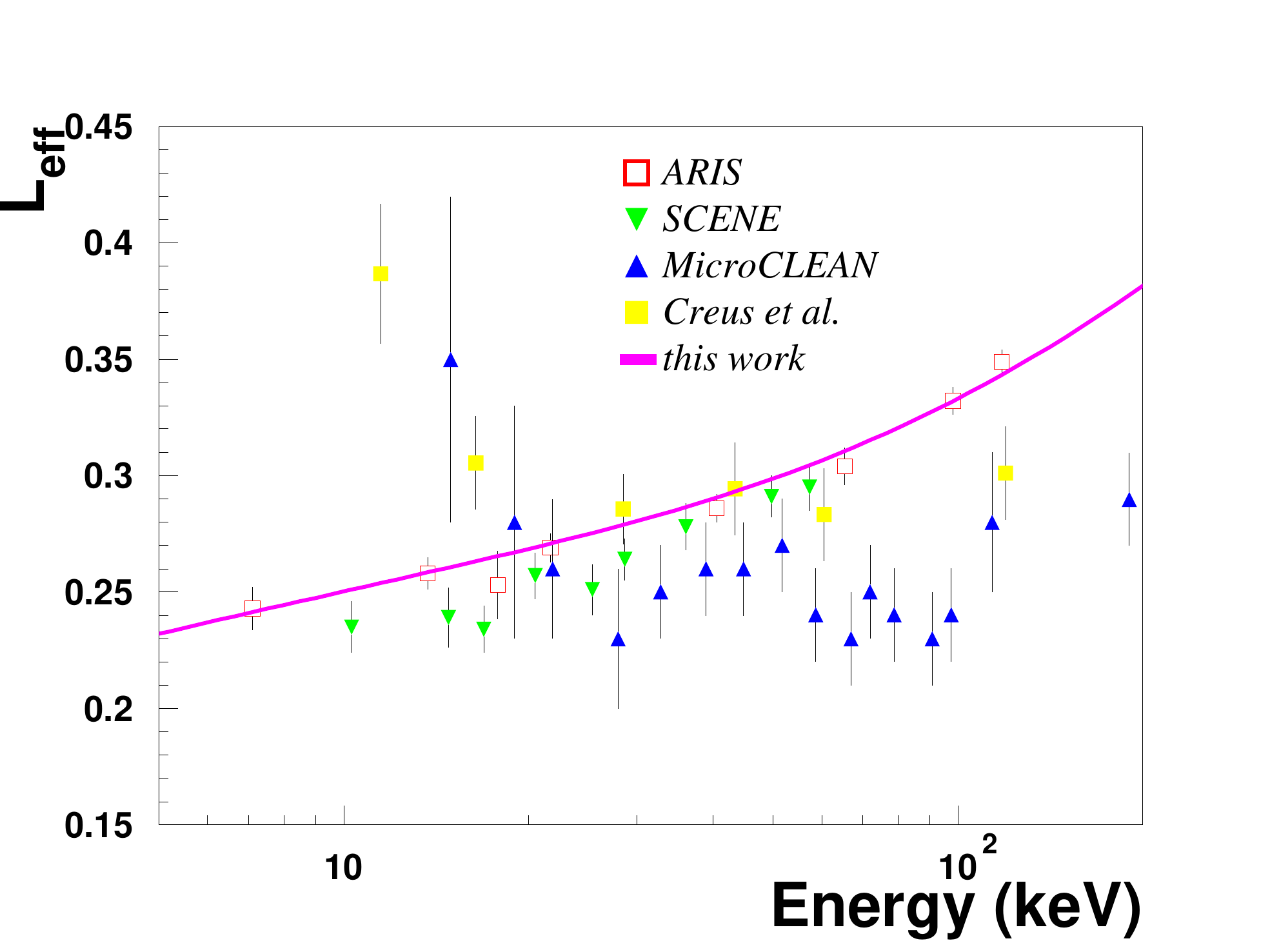}
	\caption{\label{fig:all-data-model} Comparison of the model  prediction for $\mathcal{L}_{eff}$ (magenta line) with data taken from \cite{PhysRevD.97.112005}, \cite{PhysRevD.91.092007}, \cite{PhysRevC.85.065811-microclean}, \cite{Creus_2015}.}  
\end{figure}

The result of the fit for $\mathcal{L}_{eff}$ is shown in figure \ref{fig:leff_aris}. 
The model prediction, together with the majority of available data is shown in figure \ref{fig:all-data-model}.

\section{Discussion}
\label{sec:discussion}
The triplet-triplet quenching reaction \ref{eq:quenching_self} has enough energy to ionize one of the two interacting triplet states, since the band gap of LAr is about 14.2~eV \cite{Chepel_2013}. The model assumes that the ionized molecule and the electron recombine in a triplet state. This is a good assumption if the spin relaxation time of the ion-electron system is long compared to the recombination time. The possibility that the ion-electron system recombine in a singlet state has been tested by explicitly including it in the model and a very small contribution, at the level of percent, has been found and it has been neglected.\\
The triplet-ion quenching reaction \ref{eq:quenching_ions} has enough energy to dissociate the Ar$2^+$ ion, whose binding energy is about 1.3~eV \cite{WEITZEL2002175}, and the model assumes that the Ar$+$ and Ar recombine and the density of Ar$_2^+$ states does not change.\\   
An important hypothesis of the model is to neglect the diffusion terms in equations \ref{eq:N_3_t_X} and \ref{eq:N+_t_x}. In the case of low LET particles this is well justified by the transversal distribution  (with respect to particle direction) of the density of deposited energy which is grater than hundreds of nm, while the diffusion constant of Ar$_2^*$ should be of the order of 10$^{-6}$~cm$^2$/s or less \cite{lar-self-diffusion-cini}. This assumption is also consistent with the results obtained for nuclear recoils, since the volume inside which the electronic excitation energy is found to be released is compatible with the simple diffusion of free excitons and holes before self-trapping (see Sec. \ref{sec:quenching-nuclear-recoils})  and the contribution of Ar$_2^*$ diffusion should be negligible on the time scale of LAr scintillation emission.\\
The model intorduces a possible additional quenching mechanism of the scintillation light in LAr and poses a serious question about its absolute photon yield.
This does not necessarily means that the photon yield (after the quenching) is different from what is widely assumed, since it has never been directly measured. It is believed that the ideal photon yield (51,000 photons/MeV \cite{Doke:1990rza}), estimated on the basis of the value of W$_{el}$ and N$_{ex}$/N$i$, is reached by relativistic heavy ions and then scaled for the other particles and energies according to the experimental results of relative measurements.
This point should be experimentally addressed since the technology of light detection is mature enough to allow precision measurements of the photon yield. This and other measurements and characterizations of the LAr scintillation light properties would be extremely desirable in view of the design of the next generation experiments for neutrino and Dark Matter direct detection.

\section{Conclusions}
This work describes a model for the production of LAr scintillation light which takes into account the quenching of Ar$_2^*$ through self interactions and interactions with Ar$_2^+$ ions. It allows to justify two processes which could not be explained otherwise: the dependence of the slow scintillation decay time from the intensity of an external electric field and the increase of the photon yield of xenon doped liquid argon, both for low LET particle interactions. It is possible to make an accurate prediction of the time profile of the scintillation pulse where the dependence on the electric field and on the density of the deposited energy is explicit. A simultaneous fit to experimental average waveforms of electron and nuclear recoil events  allows to constrains some of the most relevant parameters of the model such as the unquenched decay time of the slow scintillation component, $\tau_3$, which results to be around 2100~nsec and the relative abundance of singlet and triplet states for electron and nuclear recoils.\\ 
Knowing the shape of the scintillation pulse makes it possible to analytically calculate the relative abundance of slow component, F$_{prompt}$, which is often used as a pulse shape discrimination parameter.
The expressions of F$_{prompt}$ for electron and nuclear recoils  need to be fitted to the available data to extract some of the parameters which are not precisely known, but the overall behavior is well reproduced.\\
The model allows to predict the shape of the relative scintillation efficiency for nuclear recoils in LAr, $\mathcal{L}_{eff}$, and it has been shown that it reproduces closely the experimental data reported by the ARIS Collaboration \cite{PhysRevD.97.112005}.\\
LAr is a powerful medium for particle detection which is being widely used in many fields of fundamental particle physics. Deepening the knowledge of its properties can greatly benefit the design of the next generation of detectors.

\begin{acknowledgments}
	The author thanks the members of the {\it HdM} team, which is still, virtually, alive: Flavio Cavanna, Nicola Canci, Roberto Acciarri and Andrzej Szelc and which is the source of some of the data used in this work. The author warmly thanks Ana Machado for her suggestions and for the continuous discussions which helped the analyses presented here and Paola Tomassini who inspired and guided this work since the beginning.
	
\end{acknowledgments}

\bibliography{bibliography}

\end{document}